\documentclass[prl,twocolumn,aps]{revtex4-1}
\usepackage{bm}
\usepackage{graphicx}
\newcommand{\B}[1]{{\bm{#1}}}
\newcommand{\C}[1]{{\mathcal{#1}}}
\usepackage[latin1]{inputenc}
\newcommand{\beq}{\begin{equation}}
\newcommand{\eeq}{\end{equation}}
\newcommand{\bea}{\begin{eqnarray}}
\newcommand{\eea}{\end{eqnarray}}
\newcommand{\pa}{\partial}
\newcommand{\tr}{\hbox{tr}}
\usepackage{hyperref}

\pdfoutput=1

\begin{document}
\pdfoutput=1
\title{Fracture Toughness of Metallic Glasses: Ductile-to-Brittle Transition?}

\author{Chris H. Rycroft$^{1,2}$ and Eran Bouchbinder$^3$}
\affiliation{$^1$ Department of Mathematics, University of California, Berkeley, CA 94720, United States\\
$^2$ Department of Mathematics, Lawrence Berkeley Laboratory, Berkeley, CA 94720, United States\\
$^3$ Chemical Physics Department, Weizmann Institute of Science, Rehovot 76100, Israel}


\begin{abstract}
Quantitative understanding of the fracture toughness of metallic glasses, including the associated ductile-to-brittle transitions, is not yet available. Here we use a simple model of plastic deformation in glasses, coupled to an advanced Eulerian level set formulation for solving complex free boundary problems, to calculate the fracture toughness of metallic glasses as a function of the degree of structural relaxation corresponding to different annealing times near the glass temperature. Our main result indicates the existence of an elasto-plastic crack tip instability for sufficiently relaxed glasses, resulting in a marked drop in the toughness, which we interpret as a ductile-to-brittle transition similar to experimental observations.
\end{abstract}
\maketitle

The mechanical properties of glassy materials still pose challenges of great scientific and technological importance.
One such fundamental property is the fracture toughness -- the ability of a material to resist failure in the presence of a crack \cite{99Bro}. Theoretically predicting the fracture toughness of materials, which is lacking in general, is a particularly pressing problem in the context of metallic glasses. Metallic glasses constitute a promising new class of materials, possessing superior properties, whose usage in structural applications is severely limited by their relatively low fracture toughness \cite{BMG_Greer, BMG, BMGa, BMGb, Review2007, Review2008, Review2010, Toughness_review2010}.

Recent observations demonstrated a marked drop in the fracture toughness of metallic glasses as a function of composition and degree of structural relaxation (controlled through annealing near the glass temperature $T_g$) \cite{01Lewandowski, 05LWG, Review2007, Toughness_review2010}. The drop in the toughness, which is commonly correlated with Poisson's ratio \cite{01Lewandowski, 05LWG,Toughness_review2010}, is interpreted as a kind of ductile-to-brittle transition \cite{05LWG, Toughness_review2010}. To the best of our knowledge, no basic theoretical understanding of this important observation is currently available.

In this Letter we calculate the fracture toughness of metallic glasses based on the low-temperature Shear-Transformation-Zone (STZ) model, using an advanced Eulerian level set formulation for solving complex free boundary problems. We demonstrate the existence of an elasto-plastic crack tip instability as a function of increasing degree of structural relaxation, which results in a drop in the fracture toughness.
We propose to interpret this instability as a ductile-to-brittle transition similar to the one observed experimentally.

The Shear-Transformation-Zone (STZ) model of amorphous plasticity \cite{98FL, JSL04, BLP-07-I, 11FL} has recently been shown to emerge within a systematic formulation of non-equilibrium thermodynamics \cite{BLII-09, BLIII-09} and to capture a wide range of glassy deformation phenomena \cite{11FL, BLP-07-II, Manning, 08JSL, 11BLa, 12rycroft}. Its main advantage in the present context is that it offers a way to quantify the degree of structural relaxation and the deformation-driven evolution of structural disorder. Our goal here is to use the STZ model in a way that goes beyond previous analyses; rather than fixing the model parameters to quantitatively describe a given phenomenon, we treat it as a predictive model where its parameters are estimated from independent sources and another phenomenon -- crack initiation -- is studied.

We focus here on a simple version of the STZ model, retaining only salient physical ingredients. As we are interested in the fracture toughness of metallic glasses at temperatures well below the glass temperature $T_g$, we neglect all spontaneous, non-driven, relaxation processes and set the plastic rate of deformation $\B D^{pl}$ to zero for stresses below the shear yield stress $s_y$. For $\bar{s}\!\ge\! s_y$ we have
\begin{eqnarray}
\label{simple_STZ}
\B D^{pl}(\B s, T, \chi) \!=\! \tau_0^{-1} \Lambda(\chi)\,\C C(\bar{s},T)\,\left[1 - s_y/\bar{s} \right] \B s/\bar{s}\ ,
\end{eqnarray}
where $\B s\!=\!\B\sigma \!-\! \case{1}{3}\tr\B\sigma \,\B 1$ is the deviatoric stress tensor ($\B \sigma$ is the Cauchy stress) and $\sqrt{2}\,\bar{s}\!\equiv\!\sqrt{s_{ij}s_{ij}}$ \cite{BLP-07-I, 11FL}. $\B D^{pl}$ is expressed as a product of physically meaningful terms. $\tau_0^{-1}$ is a molecular vibration rate. $\Lambda(\chi)$ is the probability to find a structural fluctuation that is particularly susceptible to shear-driven rearrangements -- an STZ. It is a function of an effective disorder temperature $\chi$, to be discussed below. $\C C(\bar{s},T)$ quantifies the (dimensionless) rate in which STZs actually undergo shear transformations as a function of stress and temperature. The last terms represent deformation-induced anisotropy (``back stress'') and also make the whole expression tensorially consistent.

The effective temperature $\chi$ characterizes the out-of-equilibrium structural degrees of freedom of a glass \cite{BLII-09}. It satisfies an effective heat equation of the form \cite{JSL04, BLII-09}
\begin{eqnarray}
\label{chi}
\tau_0 \dot\chi = \Gamma(\bar{s},\chi)\,(\chi_\infty - \chi) \ ,
\end{eqnarray}
where again spontaneous thermally activated relaxation is excluded. $\chi_\infty$ is the steady state value of $\chi$ and $\Gamma(\bar{s},\chi)$ is a dimensionless strength of mechanically-generated noise that tends to rejuvenate the glass (when $\chi\!<\!\chi_\infty$). The very same theoretical framework predicts that $\Lambda(\chi)$ in Eq. (\ref{simple_STZ}) is given by a generalized Boltzmann factor, $\Lambda(\chi)=\exp{\left(-e_z/k_B\chi\right)}$, where $e_z$ is a typical STZ formation energy. It is this physical description of structural disorder that makes the STZ model most suitable for studying the fracture toughness as a function of the degree of structural relaxation.

To complete the presentation of the model in Eqs. (\ref{simple_STZ})-(\ref{chi}), we need to specify explicit forms for $\C C(\bar{s},T)$ and $\Gamma(\bar{s},\chi)$. The latter has been proposed to be proportional to the rate of plastic work $D^{pl}_{ij}s_{ij}$ \cite{11FL}, i.e. $\Gamma(\bar{s},\chi) = \tau_0 D^{pl}_{ij}s_{ij}/s_y$.
$\C C(\bar{s},T)\!\equiv\!\case{1}{2}\left[\C R(\bar{s},T)\!+\!\C R(-\bar{s},T) \right]$ is the average of forward and backward STZ transition rates $\C R(\pm\bar{s},T)\!=\!\exp\left(-\frac{\Delta\,\mp\,\Omega\,\epsilon_0\,\bar{s}}{k_B T}\right)$, which we assume to follow a linearly stress-biased thermal activation process. Here $\Delta$ is the typical energy activation barrier, $\Omega$ is the typical activation volume and $\epsilon_0$ is the typical local strain at the transition \cite{77Spa, Review2007}.
In the presence of the high stresses near a tip of a crack, $\Omega\,\epsilon_0\bar{s}$ may become larger than $\Delta$, in which case we assume
the exponential thermal activation form crosses over to a much weaker dependence associated with a linear, non-activated, dissipative mechanism \cite{08JSL}. Hence,
\begin{eqnarray}
\label{Cal_C}
\C C(\bar{s},T) \!=\! \left\{\begin{array}{ll}
\!\!e^{-\Delta/k_B T}
\cosh\left[\Omega\,\epsilon_0\,\bar{s}/k_B T\right] &\text{for}\  \Omega\,\epsilon_0\bar{s} < \Delta \vspace{0.1cm}\\
           \!\!\Omega\,\epsilon_0\,\bar{s}/2\Delta &\text{for} \  \Omega\,\epsilon_0\bar{s} \ge \Delta \ .
\end{array}\right.
\end{eqnarray}
As $\Delta\!\gg\!k_B T$, the two expressions connect continuously (but not differentiably). The slope of the linear relation was chosen so as not to introduce additional parameters. These details do not affect the qualitative nature of the results to follow.

To proceed, we adopt an Eulerian formulation and write the total rate of deformation tensor as a sum of elastic and plastic contributions, $\B D^{tot}\!=\!\B D^{el}\!+\!\B D^{pl}$, where $\B D^{tot}\!=\!\case{1}{2}[\nabla {\B v} \!+\! \left( \nabla {\B v}\right)^T]$, $\B D^{el}\!=\!\pa_t \B \epsilon\!+\!\B v \cdot \nabla \B \epsilon \!+\! \B \epsilon \cdot \B \omega \!-\!  \B \omega \cdot \B \epsilon$ and $\B \omega\!=\! \case{1}{2}[\nabla {\B v} \!-\! \left( \nabla {\B v}\right)^T]$. The strain tensor $\B\epsilon$ is related to $\B \sigma$ through Hooke's law $\B \sigma \!=\! K\,\tr\B\epsilon\B 1+2\mu\left(\B\epsilon-\case{1}{3}\tr\B \epsilon\B 1\right)$, where $K$ and $\mu$ are the bulk and shear moduli, respectively. The velocity field $\B v(\B r,t)$, where $\B r$ is the spatial coordinate, evolves through the momentum balance equation $\rho_0\left( \pa_t \B v \!+\! \B v\!\cdot\!\nabla\B v\right)\!=\!\nabla\!\cdot\!\B\sigma$, where $\rho_0$ is the mass density (assumed constant hereafter).

\begin{figure}[here]
\centering 
\includegraphics[width=3.5in, height=0.55\textheight]{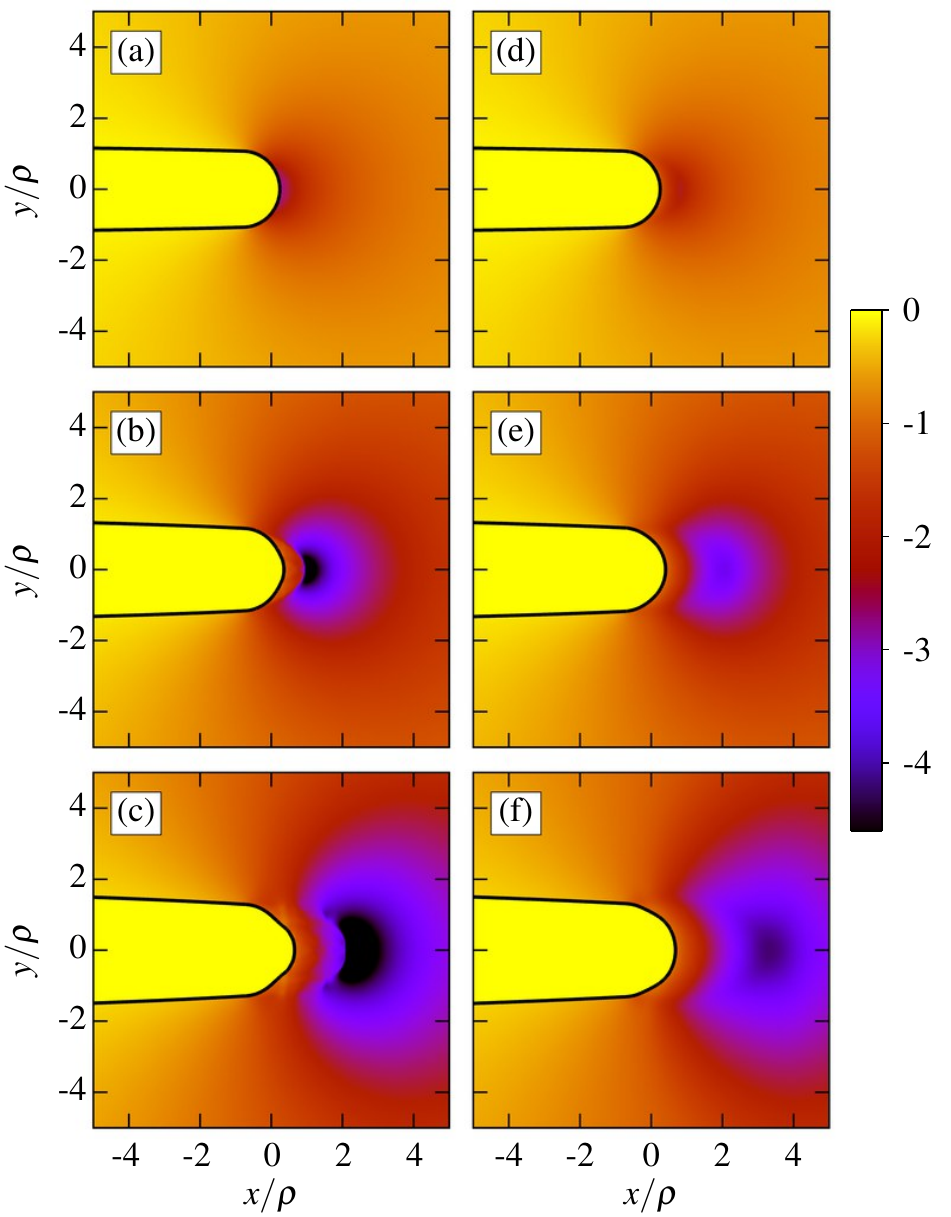}
\caption{(Color online) The normalized hydrostatic pressure field $p/s_y$ near a notch at various loading levels $K_I\!=\!20, 40, 60$MPa$\sqrt{\hbox{m}}$ (from top to bottom) for $\chi_0\!=\!600$K (panels a-c) and $\chi_0\!=\!660$K (panels d-f). A movie is available at \cite{SM}.}\label{fig1}
\end{figure}

Consider a straight notch (crack) with root radius $\rho$ (see Fig. \ref{fig1}) under plane-strain conditions. A polar coordinate system $(r,\theta)$ is set a distance $\rho/5$ behind the notch root and $\theta\!=\!0$ is the symmetry axis. We adopt a boundary layer formulation in which the following universal mode I (tensile) crack tip velocity fields are imposed on a scale much larger than $\rho$ \cite{BoundaryLayer_a, BoundaryLayer_b, TipField_2007, TipField_Anand, TipField_mixed2009, TipField_review2009}
\begin{eqnarray}
\label{Irwin}
v_x(r,\theta,t) &=& \frac{\dot{K}_I(t)}{4\mu}\!\sqrt{\frac{r}{2\pi}}\!\left[\left(5-8\nu\right)\cos\!\left(\!\frac{\theta}{2}\!\right)\!-\! \cos\!\left(\!\frac{3\theta}{2}\!\right)\!\right], \nonumber\\
v_y(r,\theta,t) &=& \frac{\dot{K}_I(t)}{4\mu}\!\sqrt{\frac{r}{2\pi}}\!\left[\left(7-8\nu\right)\sin\!\left(\!\frac{\theta}{2}\!\right)\!-\! \sin\!\left(\!\frac{3\theta}{2}\!\right)\!\right]\!,
\end{eqnarray}
where $K_I(t)$ is the mode I stress intensity factor and $\nu$ is Poisson's ratio \cite{99Bro}. The main advantage of this approach is that the stress intensity factor uniquely couples the inner scales near the tip to the outer scales and hence can be controlled independently without solving the global crack problem \cite{99Bro}.

The linear elastic fracture toughness is the critical value of the stress intensity factor, $K_{I\!c}$, at which the crack initiates and global failure occurs. There is ample experimental and numerical evidence that metallic glasses under tension fail locally near crack tips by the nucleation of voids \cite{Falk_DBT, Flores, void_craters, void_Bouchaud, void_Gao}. We interpret this at the continuum level (atomistic aspects might be also relevant \cite{Samwer2011}) as a local cavitation instability initiating at a structural fluctuation when the hydrostatic tension $\case{1}{3}\tr\B\sigma$ exceeds a threshold, which for non-hardening materials is estimated as \cite{91HHT}
\begin{equation}
\label{cavitation}
\sigma_c \simeq
2s_y\left(1+\log{\left[2\,E/(3\sqrt{3}\,s_y)
\right]} \right)/\sqrt{3} \ ,
\end{equation}
where $E$ is Young's modulus and $s_y/E \!\ll\! 1$.

The model parameters for Vitreloy 1, a widely studied metallic glass for which the annealing time dependence of the fracture toughness was measured \cite{01Lewandowski, 05LWG}, are estimated from independent sources. we set $\mu\!=\!37$GPa, $\nu\!=\!0.35$, $\rho_0\!\simeq\!6$g/cm$^{3}$ and $s_y\!\simeq\!0.85$GPa \cite{BMG, Review2007}. The basic vibrational timescale is $\tau_0\!\simeq\!10^{-13}$s. The activation volume of an STZ was estimated to be $\Omega\!\simeq\!1000{\AA}^3$ \cite{Barrier} and typically $\epsilon_0\!\simeq\!0.1$ \cite{Review2007}, hence $\Omega\epsilon_0\!\simeq\!100{\AA}^3$. The typical activation barrier is of the order of $1$eV; we set $\Delta\!=\!0.7$eV \cite{Barrier}. The STZ formation energy should be somewhat larger than $\Delta$ and we choose $e_z\!=\!1.8$eV. Finally, the steady state value of the effective temperature is expected to be between $T_g\!=\!623$K and the melting temperature $T_m\!\simeq\!1000$K. Previous works suggest $\chi_\infty\!\simeq\!900$K \cite{08JSL}. We set $T\!=\!400$K, well below $T_g$.

We set $\dot{K}_I\!=\!10$MPa$\sqrt{\hbox{m}}$\,s$^{-1}$ and $\rho\!=\!65\mu$m \cite{01Lewandowski, 05LWG}. A key parameter is the initial value of the effective temperature, $\chi(\B r,t\!=\!0)\!\equiv\!\chi_0$. In \cite{01Lewandowski, 05LWG}, Vitreloy 1 was annealed for different times at $T_g$ and the fracture toughness dropped by an order of magnitude, from $K_{I\!c}\!\simeq\!85$MPa$\sqrt{\hbox{m}}$ for the as-cast samples to $K_{I\!c}\!\simeq\!8.5$MPa$\sqrt{\hbox{m}}$ for the $12$ hours annealed samples. Within the model, we represent the effect of increasing annealing times by decreasing values of the initial effective temperature, and focus on the range $\chi_0\!=\!600\!-\!660$K. All other parameters remained fixed.

We numerically solved the equations for $\B\sigma$, $\B v$ and $\chi$ using the recently proposed Eulerian finite-difference simulation framework, where free boundaries are implicitly tracked by the level set method \cite{sethian}. A key advantage of this method is its ability to naturally handle topological changes, such as those involved in material failure. The combination of finite-difference and level set methods provides a flexible platform to study complex physical phenomena such as crack initiation and propagation.

The widely separated timescales of elastic and plastic deformations make our equations stiff. In \cite{12rycroft}, an explicit update procedure, in which the timestep was chosen to be small enough to resolve elastic waves, was employed. It would be prohibitively computationally expensive to access physically relevant timescales using this procedure. We therefore constructed a new numerical scheme in which $\B\sigma$ and $\chi$ are explicitly updated, but $\B v$ is solved for implicitly using quasi-static force balance $\nabla\!\cdot\!\B\sigma\!=\!0$. Details of the quasi-static scheme and its verification will be given elsewhere. Here we just stress that this scheme allows us to use physically realistic loading rates and to dynamically switch to the explicit scheme when rapid failure initiates. The calculations presented here employed a $-20 \!<\!x/\rho,\,y/\rho\!<\! 20$ domain, using a $1025 \!\times\! 1025$ grid. Increasing grid resolution and/or domain size did not significantly affect the results.

In Fig. \ref{fig1} we plot a sequence of three snapshots of the hydrostatic pressure field $p(\B r,t)\!=\!-\case{1}{3}\tr\B\sigma$ for $\chi_0\!=\!600$K (more relaxed) and $\chi_0\!=\!660$K (less relaxed), taken at the same value of $K_I$. The two sequences seem to exhibit a similar qualitative behavior in which $p$ attains a minimum {\em ahead} of the notch root at a distance that increases with $K_I$ \cite{TipField_2007, TipField_Anand, TipField_mixed2009, TipField_review2009}. There are, however, marked quantitative differences; the lower $\chi_0$ exhibits a significantly smaller minimum (accompanied by a sharp spatial variation) and the local notch root radius of curvature decreases, suggesting the onset of a localization process.

\begin{figure}[here]
\centering 
\includegraphics[width=3.5in]{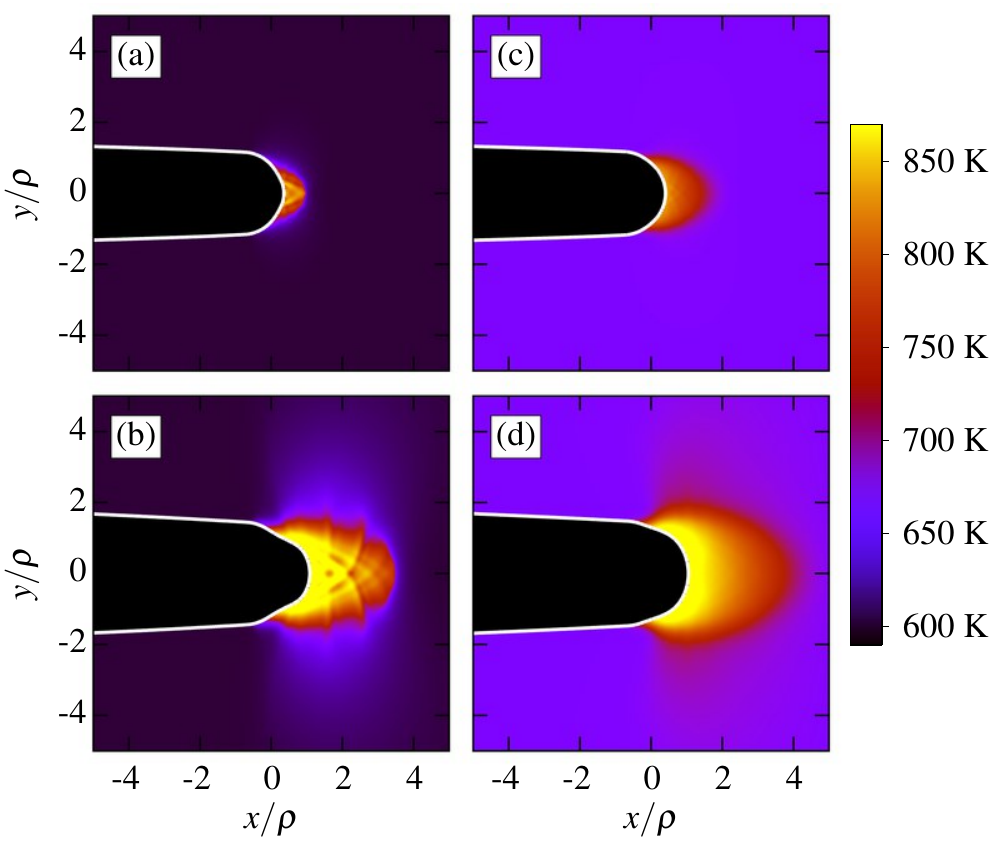}
\caption{(Color online) The effective temperature field $\chi$ at $K_I\!=\!40$MPa$\sqrt{\hbox{m}}$ (top) and $K_I\!=\!80$MPa$\sqrt{\hbox{m}}$ (bottom) for $\chi_0\!=\!600$K (panels a-b) and $\chi_0\!=\!660$K (panels c-d). A movie is available at \cite{SM}.}\label{fig2}
\end{figure}

To further explore the crack tip dynamics, we plot in Fig. \ref{fig2} two snapshots of the effective temperature $\chi(\B r,t)$ for each $\chi_0$. Recall that $\chi(\B r,t)$ quantifies structural disorder -- the higher $\chi$, the higher the disorder and the easier it is to flow. Both the spatial distribution of $\chi$ and the notch geometry are markedly different in the two cases. In the higher $\chi_0$ case, $\chi(\B r,t)$ is rather smoothly distributed in the near tip region and the notch undergoes continuous blunting -- its radius of curvature grows continuously and uniformly with $K_I$.

The lower $\chi_0$ case is qualitatively different. Initially, at small loads, there is little plastic deformation and $\chi$ remains nearly constant at its initial value $\chi_0$. As $K_I$ increases, plastic deformation localizes in the notch root vicinity, resulting in a more sharply and inhomogeneously distributed $\chi$, featuring small scale filamentary structures. These dynamics are strongly coupled to the notch geometry; the radius of curvature of the notch varies spatially, with a pronounced reduction near the root. It is this localization process -- an elasto-plastic crack tip instability -- that is responsible for the marked differences in the minima of $p$ in Fig. \ref{fig1}.

\begin{figure}[here]
\centering
\includegraphics[width=3.5in]{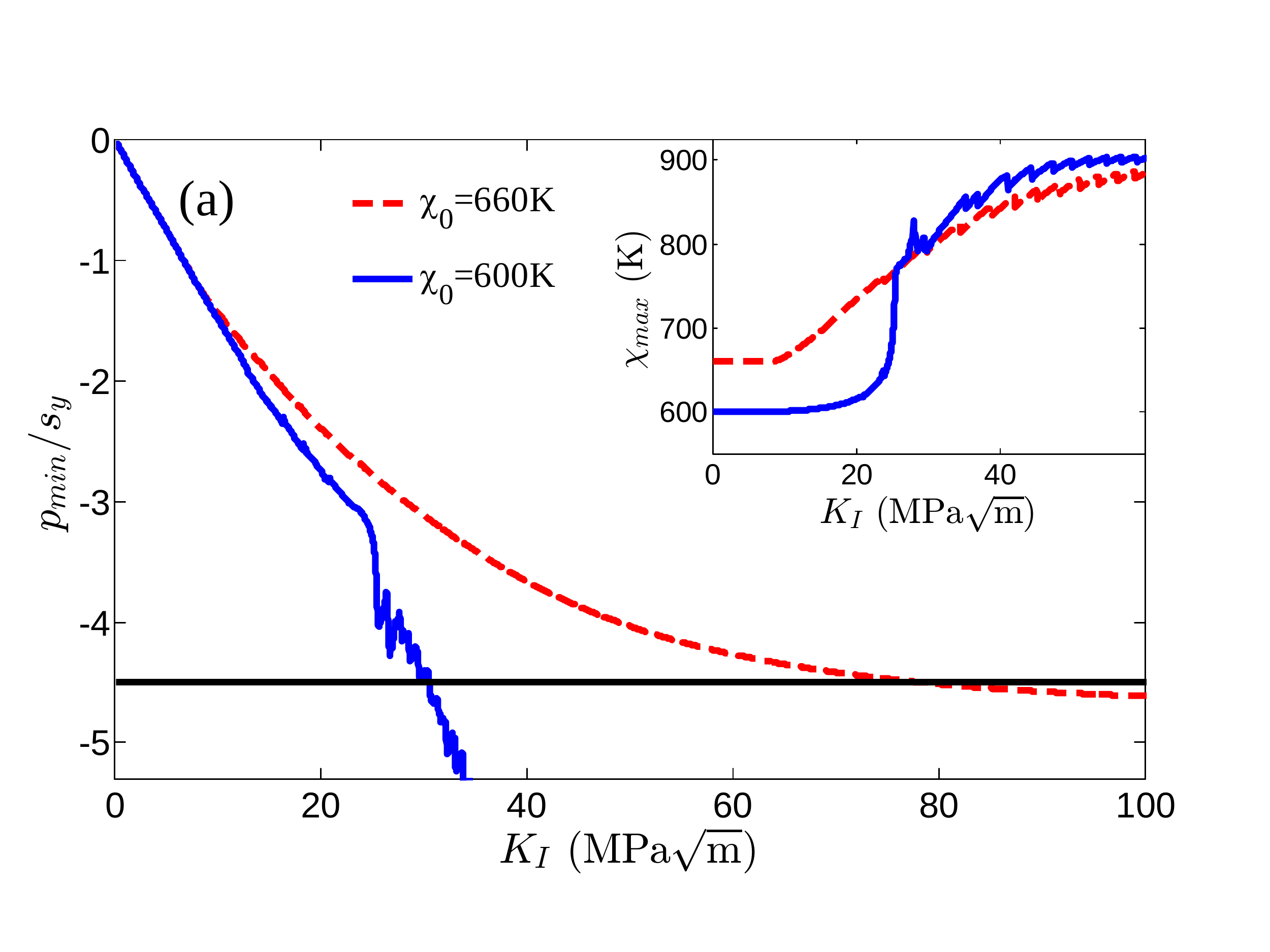}\\
\includegraphics[width=3.5in]{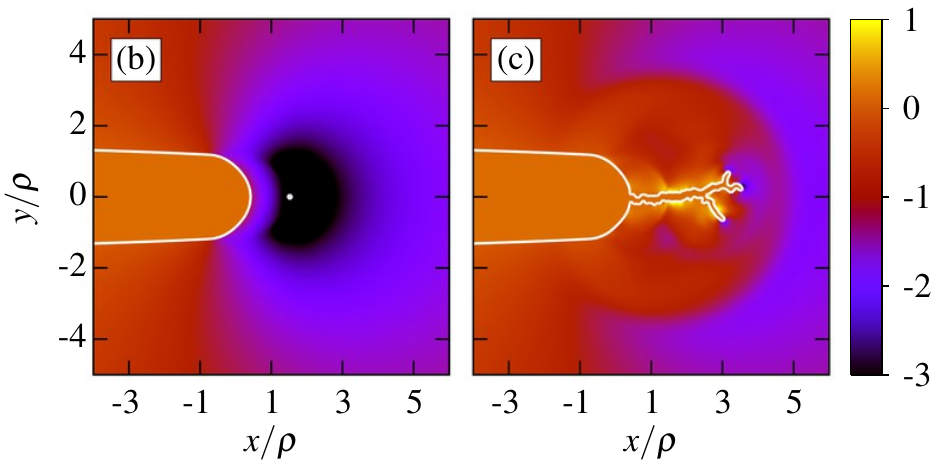}
\caption{(Color online) (a) $p_{min}/s_y$ vs. $K_I$ for $\chi_0\!=\!600$K (solid blue line) and $\chi_0\!=\!660$K (dashed red line). The horizontal line at $p_{min}/s_y\!=\!-4.5$ is the threshold for void nucleation. (inset) $\chi_{max}$ vs. $K_I$ (b) A snapshot of the system ($p/s_y$ is plotted) when a void (small white circle) nucleates. (c) The subsequent catastrophic failure. A movie is available at \cite{SM}.}\label{fig3}
\end{figure}

What are the implications of this instability for the fracture toughness? As discussed above, large $|p|$ may induce the nucleation of voids, which might lead to catastrophic failure. Therefore, we focus our attention on the minimum of the pressure $p_{min}(t)\!\equiv\!\hbox{Min}\{p(\B r,t)\}$. In Fig. \ref{fig3}a we plot $p_{min}$ vs. $K_I$ for the two $\chi_0$'s. At small $K_I$ both samples respond linear elastically (and hence identically). As $K_I$ increases, local near tip yielding occurs and the curves progressively and significantly deviate from the elastic line. Already here we observe some quantitative differences: the lower $\chi_0$ sample exhibits less plastic deformation and consequently less stress relaxation and tip blunting, resulting in more negative $p_{min}$. As $K_I$ further increases, a clear signature of the tip instability discussed above is observed, where $p_{min}$ drops abruptly for the lower $\chi_0$, while the curve for the higher one exhibits smooth and moderate variation with $K_I$.

A complementary view on the elasto-plastic nature of the instability is obtained by plotting $\chi_{max}(t)\!\equiv\!\hbox{Max}\{\chi(\B r,t)\}$, which quantifies the magnitude of plastic deformation, vs. $K_I$ in the inset of Fig. \ref{fig3}a. For $\chi_0\!=\!660$K, a linear elastic regime ($\chi_{max}\!=\!\chi_0$) is followed by a smooth and moderate increase of $\chi_{max}$ toward $\chi_\infty$. For $\chi_0\!=\!600$K the linear elastic regime is followed by an accelerated and very sharp increase of $\chi_{max}$, which is mirrored in the drop of $p_{min}$ in the main panel. To make things quantitative, we use $E/s_y\!\simeq\!85$ for Vitreloy 1 in Eq. (\ref{cavitation}) to get $\sigma_c\!\simeq\! 5 s_y$ \cite{cavitation}. In Fig. \ref{fig3}a, we chose $4.5 s_y$ as the threshold (horizontal line) for void nucleation, which suggests a large difference in the fracture toughness, $K_{I\!c}\!\simeq\!30$MPa$\sqrt{\hbox{m}}$ for $\chi_0\!=\!600$K and $K_{I\!c}\!\simeq\!80$MPa$\sqrt{\hbox{m}}$ for $\chi_0\!=\!660$K. Varying $\sigma_c$ will not change the qualitative nature of this main result, though the flatness of the $\chi_0\!=\!660$K curve suggests quantitative implications.

Does the nucleation of a void lead to catastrophic failure? i.e. can we interpret $K_I$ at which $p_{min}$ meets the threshold as the fracture toughness $K_{I\!c}$? To address this issue we take advantage of the model's dynamical nature and the numerical method's flexibility to study the post void nucleation dynamics. A void nucleation is shown in Fig. \ref{fig3}b. The subsequent dynamics, a snapshot of which is shown in panel (c), proceed through a rapid succession of void nucleations, leading to the coalescence of the initial void with the notch root and to rapid crack propagation which results in catastrophic failure. The emerging crack pattern is reminiscent of some experimental observations (random fluctuations in the void nucleation locations were introduced to avoid artificial grid effects) \cite{Notch}. In light of this catastrophic failure, we interpret the large variation in $K_I$ at which the threshold is met in Fig. \ref{fig3}a for the two different $\chi_0$'s as a kind of ductile-to-brittle transition similar to the experimental observations.

The crack tip instability, which leads to the marked drop in the fracture toughness discussed above, has both constitutive and geometric origins. The central physical question here is how efficiently a material can tame the linear elastic stress singularity, associated with the universal crack tip fields of Eqs. (\ref{Irwin}), by stress relaxation processes \cite{damage-tolerant}. Stress relaxation is mediated both by bulk plastic deformation and by the accompanying geometrical changes in the shape of the notch -- the higher the radius of curvature, the lower the stress concentration. As a glass becomes progressively more structurally relaxed (less disordered), these stress relaxation processes become progressively more limited and below some threshold a tip instability sets in.

As mentioned above, the ductile-to-brittle transition is commonly correlated with Poisson's ratio $\nu$ \cite{05LWG, Review2007, Toughness_review2010}. We suspect that this correlation might not be deep, but rather represents the fact that both the elastic and plastic responses of a glass depend on its state of disorder, quantified here by $\chi$. Hence, while there should exist a configurational equation of state of the form $\nu(\chi)$ \cite{BLP-07-II, eos} (which was neglected in our calculations), its effect on the fracture toughness is expected to be secondary compared to the strong exponential dependence of $\B D^{pl}$ on $e_z/k_B \chi_0$ through $\Lambda(\chi)$. Indeed, for our parameters $\Lambda$ drops by more than an order of magnitude when $\chi_0$ decreases from $660$K to $600$K.

The typical fracture toughness values that emerge from our calculations seem to be in the right ballpark, without fine-tuning the model's parameters. We would not, however, take this to imply that the present model has {\em quantitative} predictive powers as there are still uncertainties about the details of the model (e.g. the form of the rate factor) and the values of the parameters. On the other hand, we do advocate the view that the model can be used to {\em qualitatively} predict new phenomena, such as the crack tip instability discussed above.

E.B. acknowledges support from the Minerva Foundation with funding from the
Federal German Ministry for Education and Research, the Harold
Perlman Family Foundation and the William Z. and Eda Bess Novick
Young Scientist Fund. C.H.R. was supported by the Director, Office of Science, Computational and Technology Research, U.S. Department of Energy under contract number DE-AC02-05CH11231.

\end{document}